\documentclass{article}
\usepackage{amscd}
\usepackage{stmaryrd}
\usepackage{mathrsfs}
\usepackage{amsmath}
\usepackage{amssymb}
\usepackage{amsfonts}
\usepackage{graphicx}
\newcommand{\pa}{\partial}

\newcommand{\la}{\lambda}

\newcommand{\f}{\frac}
\newcommand{\ti}{\tilde}

\newtheorem{theorem}{Theorem}[section]
\newtheorem{proposition}[theorem]{Proposition}

\newtheorem{remark}[theorem]{Remark}
\newtheorem{example}[theorem]{Example}

\numberwithin{equation}{section}

\begin{document}
\title{{On a Novel Class of Integrable ODEs}\\
{Related to the Painlev\'e Equations}}
\author{
{A. S. Fokas${}^a$\footnote{T.Fokas@damtp.cam.ac.uk}~~~and~~D. Yang${}^{a,}{}^{b}$\footnote{yangd04@mails.tsinghua.edu.cn}}\\
\\
{\small ${}^a$ Department of Applied Mathematics and Theoretical Physics,}\\
{\small University of Cambridge, Cambridge CB3 0WA, UK}\\
\\
{\small ${}^b$ Department of Mathematical Sciences, Tsinghua
University,}\\
{\small Beijing 100084, P. R. China}\\
}
\maketitle
\begin{center}
\small \textit{This paper is dedicated to Professor T. Bountis on
the occasion of his 60th birthday with appreciation of his important
contributions to ``Nonlinear Science''.}
\end{center}
\begin{abstract}
One of the authors has recently introduced the concept of conjugate
Hamiltonian systems: the solution of the equation $h=H(p,q,t),$
where $H$ is a given Hamiltonian containing $t$ explicitly, yields
the function $t=T(p,q,h)$, which defines a new Hamiltonian system
with Hamiltonian $T$ and independent variable $h.$ By employing this
construction and by using the fact that the classical Painlev\'e
equations are Hamiltonian systems, it is straightforward to
associate with each Painlev\'e equation two new integrable ODEs.
Here, we investigate the conjugate Painlev\'e II equations. In
particular, for these novel integrable ODEs, we present a Lax pair formulation, as well as a class of implicit
solutions. We also construct conjugate equations associated with Painlev\'e I and
Painlev\'e IV equations.
\end{abstract}


\section{Introduction}
The mathematical and physical significance of Painlev\'e equations \cite{INCE}
is well established \cite{ASF5}. In particular, regarding Painlev\'e II, we note the following:
\begin{enumerate}
\item[(a)] A new method for solving its initial value problem, the
so-called isomonodromy method, was introduced in \cite{Flasch}; this
method was imbedded within the framework of the Riemann-Hilbert
formalism in \cite{ASF4}. Rigorous aspects of this formalism,
including the proof that the solution possesses the so-called
Painlev\'e property, were discussed in Fokas and Zhou
\cite{ASF_Zhou}.
\item[(b)] There exist several Lax pairs for
Painlev\'e II, including those presented in \cite{Flasch}, in Jimbo
and Miwa \cite{JM}, and in Harnad, Tracy and Widom \cite{HTW}.
\item[(c)] Painlev\'e II is a Hamiltonian system
\cite{Flasch}\cite{Okamoto}\cite{M.Noumi}.
\item[(d)] It is possible
to construct certain particular explicit solutions of Painlev\'e II
using certain ``B\"acklund transformations'' \cite{Yablonskii}
\cite{M.Noumi} and \cite{ASF2}.
\end{enumerate}

The concept of conjugate Hamiltonian systems is introduced in
\cite{Yang}: The solution of the equation $h=H(p,q,t),$ where $H$ is
a given Hamiltonian which contains $t$ explicitly, yields the
function $t=T(p,q,h)$. The Hamiltonian system with Hamiltonian $T$
and independent variable $h$ is called \textit{conjugate} to the
Hamiltonian system with Hamiltonian $H$. The conjugate Hamiltonian
system has the following properties:
\begin{enumerate}
\item If $p=p(t),~q=q(t)$ is a solution of the Hamiltonian system with
Hamiltonian $H$, then $p=p(t(h)),q=q(t(h))$\footnote{
We sometimes write $p(h),q(h)$ instead of $p(t(h)),q(t(h))$.}, is a solution
of the conjugate Hamiltonian system, where $t=t(h)$ is the so-called
$t$-function, the inverse function of the $h-$function\footnote{We
recall that the $h-$function, $h=h(t)$, is defined by
$h(t)=H(p(t),q(t),t))$.}.

\item A first integral of a Hamiltonian system, also provides a first integral of
the associated conjugate Hamiltonian system.
\end{enumerate}

The classical Painlev\'e equations are Hamiltonian systems, thus we
can associate with each Painlev\'e equation a conjugate Hamiltonian
system. The gauge freedom of a Hamiltonian implies that
we can in fact associate an \emph{infinite} family of integrable
second-order nonlinear ODEs with a given Painlev\'e equation.
Furthermore, by utilising the gauge freedom of the conjugate
Hamiltonian, we can associate with any of the conjugate ODEs
constructed, another infinite family of integrable ODEs, etc.
Here, we only present the ODEs with the
simplest form.

This paper is organized as follows: In section 2, we construct the
conjugate Painlev\'e II and also derive an associated Lax pair. In
section 3, we construct a class of explicit solutions of the
conjugate Painlev\'e II. In section 4, we derive the conjugate ODEs
corresponding to Painlev\'e I and IV. In section 5, we prove a general theorem
for constructing Lax pairs for conjugate Painlev\'e equations. In
section 6, we discuss further these results.

\section{The conjugate Painlev\'e II equation}
Let $P_{I\!I}$ denote the second Painlev\'e equation, namely
\begin{equation}
\label{P2_q}  P_{I\!I}:~~~\f{d^2q}{dt^2}=2q^3+tq+b-\f{1}{2},~~~~t,q\in\mathbb{C},
\end{equation}
where $b$ is an arbitrary complex constant. $P_{I\!I}$ possesses the
Hamiltonian $H$, where
\begin{equation}\label{P2_H}
H(p,q,t)=\f{1}{2}p^2-\Big(q^2+\f{t}{2}\Big)p-bq.
\end{equation}
Indeed, Hamilton's equations associated with $H$ are
\begin{subequations}\label{Ham_P2}
 \begin{align}
   \label{Ham_P2_a} &\f{dp}{dt}=-\f{\pa H}{\pa q}=2pq+b,\\
   \label{Ham_P2_b} &\f{dq}{dt}=\f{\pa H}{\pa p}=p-q^2-\f{t}{2}.
 \end{align}
\end{subequations}
Eliminating from equations \eqref{Ham_P2} the variable $p$ we find
$P_{I\!I}:$
$$\f{d^2q}{dt^2}=2pq+b-2q\Big(p-q^2-\f{t}{2}\Big)-\f{1}{2},$$
which is equation \eqref{P2_q}.

If we eliminate from equations \eqref{Ham_P2} the variable $q$, we
find an other second order integrable ODE, which appears in the list
of Ince \cite{INCE} as $XXXIV$, and which we denote by
$\widetilde{P_{I\!I}}$ \footnote{Either $\widetilde{P_{I\!I}}$ or
$P_{I\!I}$ could have been chosen as the second Painlev\'e equation. The
historical choice of $P_{I\!I}$ is due to Painlev\'e himself
\cite{INCE}.}:
\begin{equation}
\label{P2_p}
\widetilde{P_{I\!I}}:~~\f{d^2 p}{dt^2}=\f{1}{2p}\Big(\f{dp}{dt}\Big)^2-\f{b^2}{2p}+2p^2-tp,~~~~t,p\in\mathbb{C}.
\end{equation}
Indeed,
\begin{equation*}
 \begin{split}
  \f{d^2 p}{dt^2}&=2q(2pq+b)+2p\Big(p-q^2-\f{t}{2}\Big)\\
                 &=2p^2-tp+2pq^2+2bq.
 \end{split}
\end{equation*}
Replacing in this equation $q$ from equation \eqref{Ham_P2_a}, we
find equation \eqref{P2_p}.
\begin{remark}\label{solution corres}
It has been shown in \cite{ASF2} that there exists a one-to-one
correspondence between solutions of $P_{I\!I}$ and
$\widetilde{P_{I\!I}}$. This result follows directly from
their Hamiltonian structure \eqref{Ham_P2}.
\end{remark}
\paragraph{Notations} Prime, `` ${}'$ '', denotes derivative with respect to $h$.

\begin{proposition}
The conjugate equations of $P_{I\!I}$ and of $\widetilde{P_{I\!I}}$,
i.e., the conjugate equations of equations \eqref{P2_q} and
\eqref{P2_p}, are the following ODEs:
\begin{subequations} \label{C_P2}
\begin{align}
\label{P2_C_q}CP_{I\!I}:~~&\f{d^2 q}{dh^2}=(q'+1)\Big(\f{1-2b-bq'}{h+bq}+8q\Big(\f{-q'-1}{2h+2bq}\Big)^{\f{1}{2}}\Big),~~h,q\in\mathbb{C},\\
\label{P2_C_p}\widetilde{CP_{I\!I}}:~~&\f{d^2 p}{dh^2}=4+\f{8h}{p^2}-\f{4b^2}{p^3}, ~~~h,p\in \mathbb{C}.
\end{align}
\end{subequations}
Equations \eqref{C_P2} possess the Hamiltonian function $T$, where
\begin{equation}
\label{CH2_T} T(p,q,h)=(p-2q^2)-\f{2h+2bq}{p}.
\end{equation}
Moreover, equations \eqref{C_P2} admit the following Lax pair:
\begin{subequations}\label{Lax_CP2}
 \begin{align}
\f{\pa \psi}{\pa \la}&=\left\{\f{1}{2\la}\left(
\begin{array}{llcl}
b  & 0\\
-p & -b\\
\end{array}
\right)+\left(
\begin{array}{llcl}
q & \f{1}{p}(2h+2bq)\\
\f{1}{2} & -q\\
\end{array}
\right)+\la\left(
\begin{array}{llcl}
 0 & 1\\
 0 & 0\\
\end{array}
\right)\right\}
\psi,\\
\f{\pa \psi}{\pa h}&=\left\{-\f{2}{p}\left(
\begin{array}{llcl}
-q        & 0\\
-\f{1}{2} & q\\
\end{array}
\right)-\la\f{2}{p}\left(
\begin{array}{llcl}
 0 & -1\\
 0 & 0\\
\end{array}
\right)\right\} \psi, ~~~~~~~~~\la\in\mathbb{C},
 \end{align}
\end{subequations}
where $\psi$ is a $2\times 2$ matrix-valued function of $\la$ and
$h$.
\end{proposition}

\paragraph{Proof.}
The solution of the equation
\begin{equation}
 h=\f{1}{2}p^2-(q^2+\f{t}{2})p-bq,
\end{equation}
yields
\begin{equation}
 t=T(p,q,h),
\end{equation}
where the function $T$ denotes the RHS of equation \eqref{CH2_T}. The associated Hamilton's equations are:
\begin{subequations}\label{CH2_PQ}
\begin{align}
&\f{dp}{dh}=\f{\pa T}{\pa q}=-4q-\f{2b}{p},\\
&\f{dq}{dh}=-\f{\pa T}{\pa p}=-1-\f{2h+2bq}{p^2}.
\end{align}
\end{subequations}
Eliminating from equations \eqref{CH2_PQ} the function $q$ we find,
$$\f{d^2 p}{dh^2}=-4\Big(-1-\f{2h+2bq}{p^2}\Big)+\f{2b}{p^2}\Big(-4q-\f{2b}{p}\Big),$$
which is equation \eqref{P2_C_p}. Similarly, eliminating from equations \eqref{CH2_PQ} the function $p$ we find equation \eqref{P2_C_q}.

The Lax pair \eqref{Lax_CP2} can be verified directly. Indeed, under
the assumption that $p_{\la}=0$ and $q_{\la}=0$,
i.e, the compatibility equation
$$\f{\pa^2\psi}{\pa h\pa \la}=\f{\pa^2\psi}{\pa \la \pa h},$$
is equivalent to equations \eqref{CH2_PQ}, and so is equivalent to
equations \eqref{C_P2}.

Alternatively, equations \eqref{Lax_CP2} can be derived from the following Lax pair of
$P_{I\!I}$ and $\widetilde{P_{I\!I}}$\footnote{This pair is the
so-called Harnad-Tracy-Widom pair (HTW-pair), which was first
discovered by Harnad, Tracy and Widom in \cite{HTW} and first
written out explicitly by Joshi, Kitaev and Treharne in
\cite{Kita}.}:
\begin{subequations}\label{Lax_P2_HTW}
 \begin{align}
\f{\pa \psi}{\pa \la}&=\left\{\f{1}{2\la}\left(
\begin{array}{llcl}
b  & 0\\
-p & -b\\
\end{array}
\right)+\left(
\begin{array}{llcl}
q & p-2q^2-t\\
\f{1}{2} & -q\\
\end{array}
\right)+\la\left(
\begin{array}{llcl}
 0 & 1\\
 0 & 0\\
\end{array}
\right)\right\}
\psi,\\
\f{\pa \psi}{\pa t}&=-\left\{\left(
\begin{array}{llcl}
q        & 0\\
\f{1}{2} & -q\\
\end{array}
\right)+\la\left(
\begin{array}{llcl}
 0 & 1\\
 0 & 0\\
\end{array}
\right)\right\}
\psi.
\end{align}
\end{subequations}
Indeed, the HTW-pair \eqref{Lax_P2_HTW} is a Lax pair for
Hamilton's equations \eqref{Ham_P2}  \cite{Kita}. By applying Proposition \ref{Lax pair},
it can be shown that equations \eqref{Lax_P2_HTW} imply equations \eqref{Lax_CP2}.
\begin{flushright}  $\square$ \end{flushright}

\section{Solving $CP_{I\!I}$ and $\widetilde{CP_{I\!I}}$}
The discussion in the introduction implies that starting with the well-known
special solutions of $P_{I\!I}$ and $\widetilde{P_{I\!I}}$, we can
construct special solutions for $CP_{I\!I}$ and
$\widetilde{CP_{I\!I}}$. It also implies that we can solve, at least implicitly, the general initial
problem.

\subsection{A class of special solutions}
First we recall the rational solutions of $P_{I\!I}$ and
$\widetilde{P_{I\!I}}$. There are two fundamental types of
B\"acklund transformations for $P_{I\!I}$ and $\widetilde{P_{I\!I}}$
which were derived in \cite{Yablonskii}, \cite{ASF2},
\cite{M.Noumi}. Taking into consideration Remark \ref{solution
corres}, we express these transformations for Hamilton's equations
\eqref{Ham_P2}:
\begin{enumerate}
\item[$(i)$]  Suppose that $(q(t;b),p(t;b))$ is a solution of equations \eqref{Ham_P2}
with constant $b$. Then
$$(\hat{q}(t),\hat{p}(t))=(q(t;b)+\f{b}{p(t;b)},~p(t;b))$$ is a solution
of equations \eqref{Ham_P2} with constant $-b$.
\item[$(ii)$] Suppose that $(q(t;b),p(t;b))$ is a solution for equations \eqref{Ham_P2}
with constant $b$. Then
$$(\hat{q}(t),\hat{p}(t))=(-q(t;b),~-p(t;b)+2q^2(t;b)+t)$$ is a solution
of equations \eqref{Ham_P2} with constant $1-b$.
\end{enumerate}
The transformations $(i)$ and $(ii)$ imply, respectively,
$\hat{h}(t)=h(t)$ and $\hat{h}(t)=h(t)+q(t)$.

Starting from a particular solution $q=0,p=t/2$ with $b=1/2$, and
applying the above B\"acklund transformations, we can obtain a
class of rational solutions
\cite{Yablonskii}, \cite{ASF2}, \cite{M.Noumi} for $P_{I\!I}$ and
$\widetilde{P_{I\!I}}$. For example:
\begin{enumerate}
\item[   ] $q=\f{2(t^3-2)}{t(t^3+4)}$, $p=\f{t^3+4}{2t^2},$
$h=-\f{t^2}{8}+\f{1}{t}$ with $b=-\f{3}{2}$;
\item[   ] $q=\f{1}{t}$, $p=\f{t}{2}$, $h=-\f{t^2}{8}$ with $b=-\f{1}{2}$;
\item[   ] $q=0$, $p=\f{t}{2}$, $h=-\f{t^2}{8}$ with $b=\f{1}{2}$;
\item[   ] $q=-\f{1}{t}$, $p=\f{t^3+4}{2t^2}$, $h=-\f{t^2}{8}+\f{1}{t}$ with $b=\f{3}{2}$;
\end{enumerate}

By inverting the $h-$function and substituting the resulting
$t-$function to the rational solutions for $P_{I\!I}$ and
$\widetilde{P_{I\!I}}$, we obtain the following solutions for
$CP_{I\!I}$ and $\widetilde{CP_{I\!I}}$:
\begin{itemize}
\item{
\begin{subequations}\label{CP_sol_1}
\begin{align}
&q=\f{2\Big(\Big(\f{2}{3}\Big)^{2/3}D^{1/3}-\Big(\f{2}{3}\Big)^{1/3}4hD^{-1/3}\Big)^3-4}
{\Big(\f{2}{3}\Big)^{2/3}D^{1/3}-\Big(\f{2}{3}\Big)^{1/3}4hD^{-1/3}\Big[\Big(\Big(\f{2}{3}\Big)^{2/3}D^{1/3}-\Big(\f{2}{3}\Big)^{1/3}4hD^{-1/3}\Big)^3+4\Big]},\\
&p=\f{4+\Big(\Big(\f{2}{3}\Big)^{2/3}D^{1/3}-\Big(\f{2}{3}\Big)^{1/3}4hD^{-1/3}\Big)^3}{2\Big(\Big(\f{2}{3}\Big)^{2/3}D^{1/3}
-\Big(\f{2}{3}\Big)^{1/3}4hD^{-1/3}\Big)^2},\\
&t=\Big(\f{2}{3}\Big)^{2/3}D^{1/3}-\Big(\f{2}{3}\Big)^{1/3}4hD^{-1/3}
\end{align}
\end{subequations}
with $b=-\f{3}{2}$, where $D=9+\Big(81+96h^3\Big)^{1/2}$;}
\item{ \begin{equation}q=\f{1}{2(-2h)^{1/2}}, ~~p=(-2h)^{1/2}, ~~t=2(-2h)^{1/2} \end{equation} with
$b=-\f{1}{2}$;}
\item{
\begin{equation} q=0,~~ p=(-2h)^{1/2},~~ t=2(-2h)^{1/2} \end{equation} with $b=\f{1}{2}$;
}
\item{
\begin{subequations}\label{CP_sol_4}
\begin{align}
&q=-\Big(\Big(\f{2}{3}\Big)^{2/3}D^{1/3}-\Big(\f{2}{3}\Big)^{1/3}4hD^{-1/3}\Big)^{-1},\\
&p=\f{4+\Big(\Big(\f{2}{3}\Big)^{2/3}D^{1/3}-\Big(\f{2}{3}\Big)^{1/3}4hD^{-1/3}\Big)^3}{2\Big(\Big(\f{2}{3}\Big)^{2/3}D^{1/3}
-\Big(\f{2}{3}\Big)^{1/3}4hD^{-1/3}\Big)^2},\\
&t=\Big(\f{2}{3}\Big)^{2/3}D^{1/3}-\Big(\f{2}{3}\Big)^{1/3}4hD^{-1/3}
\end{align}
\end{subequations}
with $b=\f{3}{2}$, where $D=9+\Big(81+96h^3\Big)^{1/2}$;}
\end{itemize}
The four particular solutions computed above can be verified
directly. For $|b|>3/2$, in order to compute the corresponding solutions we
need to solve polynomial equations of order higher than 4.

Using the transformations $(i)$ and $(ii)$, we find the following result:
\begin{proposition} (B\"acklund transformations)\label{B_trans}
\begin{enumerate}
\item[$(i)$]  Suppose that $(q(h;b),p(h;b))$ is a solution of equations
\eqref{CH2_PQ} with constant $b$. Then
$$(\hat{q}(h),~\hat{p}(h))=(q(h;b)+\f{b}{p(h;b)},~p(h;b))$$ is a solution of
equations \eqref{CH2_PQ} with constant $-b$.
\item[$(ii)$] Suppose that $(q(h;b),p(h;b))$ is a solution of equations \eqref{Ham_P2}
with constant $b$. Then
$$(\hat{q}(\hat{h}),~\hat{p}(\hat{h})=(-q(h;b),~\f{-2h+2bq(h;b)}{p(h;b)})$$ is a solution
of equations \eqref{Ham_P2} with constant $1-b$ and independent variable $\hat{h}$,
where $\hat{h}=h+q(h;b)$.
\end{enumerate}
\end{proposition}
\begin{remark}
The solutions \eqref{CP_sol_1} -- \eqref{CP_sol_4} can also be
generated by employing Proposition \ref{B_trans}. The main
difficulty for the explicit computation of these solutions is the
requirement of solving the equation $\hat{h}=h+q(h;b)$ for $h$ in
terms of $\hat{h}$, $h=h(\hat{h})$.
\end{remark}

\subsection{An implicit representation of the solution of the initial value problem}
We study the following initial value problem (IVP) of
$\widetilde{CP_{I\!I}}$:
\begin{subequations}
\begin{align}
&\f{d^2 p}{dh^2}=4+\f{8h}{p^2}-\f{4b^2}{p^3},\\
&p|_{h=h_0}=p_0,~~p'|_{h=h_0}=p_1.
\end{align}
\end{subequations}

$\widetilde{CP_{I\!I}}$ is equivalent to Hamilton's equations
\eqref{CH2_PQ}. Using these equations, we can find the
initial values of $q$ and $q'$ at $h=h_0:$
$$q|_{h=h_0}=-\f{b}{2p_0}-\f{p_1}{4}:=q_0,~~q'|_{h=h_0}=-1-\f{2h_0+2bp_1}{p_0^2}-\f{b^2}{p_0^3}:=q_1.$$
Thus,
$$t_0=T(p_0,q_0,h_0),~~q|_{t=t_0}=q_0,~~p|_{t=t_0}=p_0.$$

Next, from the Hamiltonian structure \eqref{Ham_P2} of $P_{I\!I}$,
we obtain the following initial values:
$$\f{dq}{dt}\Big|_{t=t_0}=2p_0q_0+b,~~~\f{dp}{dt}\Big|_{t=t_0}=p_0-q_0^2-t_0/2.$$

The IVP of $P_{I\!I}$ with initial values $q|_{t=t_0}$ and
$\f{dq}{dt}\big|_{t=t_0}$, can be solved via the isomonodromy method and yields
$$q=q(t).$$
Substituting this solution to equation \eqref{Ham_P2}, we obtain
$$p=p(t).$$ The $h-$function is obtained by
$$h(t)=H(p(t),q(t),t).$$
By the inverse function theorem, as least locally, we obtain
$$t=t(h).$$ Thus, the implicit solution of the IVP of
$\widetilde{CP_{I\!I}}$ is given by
$$p=p(t(h)).$$

The IVP for $CP_{I\!I}$ can be solved in a similar way.

\section{Conjugate equations of Painlev\'e I and IV}
Let $P_{I}$ denote the first Painlev\'e equation, namely
\begin{equation}\label{P1_q}
 P_I:~~ \f{d^2 q}{dt^2}=6q^2+t,~~~~t,q\in\mathbb{C}.\\
\end{equation}
$P_{I}$ possesses the Hamiltonian $H$, where
\begin{equation} \label{P1_H}
H(p,q,t)=\f{1}{2}p^2-2q^3-tq.
\end{equation}
Indeed, the associated Hamilton's equations are
\begin{subequations}\label{Ham_P1}
 \begin{align}
\label{Ham_P1_p}& \f{dp}{dt}=-\f{\pa H}{\pa q}=6q^2+t,\\
\label{Ham_P1_q}& \f{dq}{dt}=\f{\pa H}{\pa p}=p.
 \end{align}
\end{subequations}
Eliminating from equations \eqref{Ham_P1} the variable $p$ we find $P_I$:
$$\f{d^2 q}{dt^2}=6q^2+t.$$
Eliminating from equations \eqref{Ham_P1} the variable $q$ we find $\widetilde{P_I}$:
\begin{equation}\label{P1_p}
 \widetilde{P_I}:~~\f{d^2 p}{dt^2}=2p\Big(6\f{dp}{dt}-6t\Big)^{1/2}+1,~~~~t,p\in\mathbb{C}.
\end{equation}
Indeed,
$$\f{d^2 p}{dt^2}=12qp+1.$$
Replacing in this equation $q$ from equation \eqref{Ham_P1_p}, we find equation \eqref{P1_p}.

\begin{proposition}
The conjugate equations of equations $P_I$ and of $\widetilde{P_I}$,
i.e., the conjugate equations of equations \eqref{P1_q} and
\eqref{P1_p}, are the following ODEs:
\begin{subequations}\label{CP_1}
\begin{align}
CP_I:~~\label{CP1_q}&\f{d^2 q}{dh^2}=-\f{1}{2q}\Big(\f{dq}{dh}\Big)^2+4-\f{h}{q^3},~~~h,q\in\mathbb{C},\\
\widetilde{CP_I}:~~\label{CP1_p}&\f{d^2 p}{dh^2}=\f{2hp-p^3}{F(p',p,h)^4}+\f{1-pp'}{F(p',p,h)^2}+\f{4p}{F(p',p,h)},~~~h,p\in\mathbb{C},
\end{align}
\end{subequations}
where $F$ is a solution of the following equation $$4F^3+p'F^2+\f{1}{2}p^2-h=0,~~~h,p,F\in\mathbb{C}.$$
Equations \eqref{CP_1} possess the Hamiltonian $T$, where
\begin{equation}\label{CH1_T}
T(p,q,h)=\f{1}{2}\f{p^2}{q}-2q^2-\f{h}{q}.
\end{equation}
\end{proposition}

\paragraph{Proof.}
The solution of the equation
\begin{equation}
 h=\f{1}{2}p^2-2q^3-tq,
\end{equation}
yields
\begin{equation}
 t=T(p,q,h),
\end{equation}
where the function $T$ denotes the RHS of equation \eqref{CH1_T}. The associated Hamilton's equations are:
\begin{subequations}\label{CH1_PQ}
\begin{align}
&\f{dp}{dh}=\f{\pa T}{\pa q}=-\f{1}{2}\f{p^2}{q^2}-4q+\f{h}{q^2},\\
&\f{dq}{dh}=-\f{\pa T}{\pa p}=-\f{p}{q}.
\end{align}
\end{subequations}
Eliminating from equations \eqref{CH1_PQ} the function $p$ we find,
$$\f{d^2 q}{dh^2}=-\f{1}{2q}\Big(\f{dq}{dh}\Big)^2+4-\f{h}{q^3},$$
which is equation \eqref{CP1_q}. Similarly, eliminating from equations \eqref{CH1_PQ} the function $q$ we find equation \eqref{CP1_p}.
\begin{flushright}  $\square$ \end{flushright}

Let $P_{I\!V}$ denote the fourth Painlev\'e equation, namely
\begin{equation}\label{P4_q}
P_{I\!V}:~~\f{d^2 q}{dt^2}=\f{1}{2q}\Big(\f{d q}{dt}\Big)^2+\f{3}{2}q^3+2tq^2+\Big(\f{t^2}{2}+a_1+2a_2-1\Big)q-\f{a_1^2}{2q}, ~t,q\in\mathbb{C}.
\end{equation}
where $a_1,a_2$ are arbitrary complex constants.
$P_{I\!V}$ possesses the Hamiltonian $H$, where
\begin{equation}\label{P4_H}
H(p,q,t)=qp(p-q-t)-a_2q-a_1p.
\end{equation}
The associated Hamilton's equations are
\begin{subequations}\label{Ham_P4}
 \begin{align}
\label{Ham_P4_p}& \f{dp}{dt}=-\f{\pa H}{\pa q}=-p^2+2pq+pt+a_2,\\
\label{Ham_P4_q}& \f{dq}{dt}=\f{\pa H}{\pa p}=-q^2+2pq-qt-a_1.
 \end{align}
\end{subequations}
Eliminating from equations \eqref{Ham_P4} the variable $p$ we find $P_{I\!V}$:
$$\f{d^2 q}{dt^2}=\f{1}{2q}\Big(\f{d q}{dt}\Big)^2+\f{3}{2}q^3+2tq^2+\Big(\f{t^2}{2}+a_1+2a_2-1\Big)q-\f{a_1^2}{2q}.$$
Eliminating from equations \eqref{Ham_P4} the variable $q$ we find $\widetilde{P_{I\!V}}$:
\begin{equation}\label{P4_p}
\widetilde{P_{I\!V}}:~~\f{d^2 p}{dt^2}= \f{1}{2p}\Big(\f{d
p}{dt}\Big)^2+\f{3}{2}p^3-2tp^2+\Big(\f{t^2}{2}-2a_1-a_2+1\Big)p-\f{a_2^2}{2p},~~~t,p\in\mathbb{C}.
\end{equation}
Indeed,
$$\f{d^2 p}{dt^2}=-2p\f{dp}{dt}+2p(-q^2+2pq-qt-a_1)+2\f{dp}{dt}q+\f{dp}{dt}t+p.$$
Replacing in this equation $q$ from equation \eqref{Ham_P4_p}, we
find equation \eqref{P4_p}.
\begin{proposition}
The conjugate equations of equations $P_{I\!V}$ and of
$\widetilde{P_{I\!V}}$, i.e., the conjugate equations of equations
\eqref{P4_q} and \eqref{P4_p}, are the following ODEs:
\begin{subequations}\label{CP_4}
\begin{align}
\label{P4_C_q}
\begin{split}CP_{I\!V}:~~
\f{d^2 q}{dh^2}=\f{1+q'}{hq+a_2 q^2}\big(q-2q^2(1+q')G_1+&2(h+a_1G_1)+q'(h+2a_1G_1)\big),\\
&h,q\in\mathbb{C},
\end{split}\\
\label{P4_C_p}
\begin{split}
\widetilde{CP_{I\!V}}:~~
\f{d^2 p}{dh^2}=\f{1+p'}{hp+a_1 p^2}\big(p+2p^2(1+p')G_2+&2(h+a_2G_2)+p'(h+2a_2G_2)\big),\\
&h,p\in\mathbb{C},
\end{split}
\end{align}
\end{subequations}
where $G_1=\Big(-\f{h+a_2q}{q+qq'}\Big)^{1/2},~G_2=\Big(\f{h+a_1p}{p+pp'}\Big)^{1/2}.$

Equations \eqref{CP_4} possess the Hamiltonian $T$, where
\begin{equation}\label{CH4_T}
T(p,q,h)=p-q-\f{a_2}{p}-\f{a_1}{q}-\f{h}{pq}.
\end{equation}
\end{proposition}
\paragraph{Proof.}
The solution of the equation
\begin{equation}
 h=qp(p-q-t)-a_2q-a_1p,
\end{equation}
yields
\begin{equation}
 t=T(p,q,h),
\end{equation}
where the function $T$ denotes the RHS of equation \eqref{CH4_T}. The associated Hamilton's equations are:
\begin{subequations}\label{CH4_PQ}
\begin{align}
&p'=\f{\pa T}{\pa q}=\f{h}{q^2p}+\f{a_1}{q^2}-1,\\
&q'=-\f{\pa T}{\pa p}=-\f{h}{p^2q}-\f{a_2}{p^2}-1.
\end{align}
\end{subequations}
Eliminating from equations \eqref{CH4_PQ} the function $p$ we find \eqref{P4_C_q}.
Similarly, eliminating from equations \eqref{CH4_PQ} the function $p$ we find equation \eqref{P4_C_p}.
\begin{flushright}  $\square$ \end{flushright}

\begin{remark}
Every hamiltonian has the gauge freedom $\ti{H}=H+f(t)$, where
$f(t)$ is an arbitrary function of $t$. This implies that we can associate infinitely
many ODEs with each Painlev\'e equation. Among these ODEs, the ODEs presented here are expected to have
the simplest form.
\end{remark}

\begin{remark}
We note that conjugate Painlev\'e equations are of the form
$y''=F(y,y',t)$, where $F$ is algebraic in $y,y'$. The corresponding
conjugate Hamiltonian systems are of the form
\begin{subequations}
 \begin{align}
   p'=F_1(p,q,h),\\
   q'=F_2(p,q,h),
 \end{align}
\end{subequations}
where $F_1$ and $F_2$ are rational in $p,q.$
\end{remark}

\section{Lax pairs for conjugate equations}
The following proposition provides a method for constructing Lax pairs for conjugate Painlev\'e equations.
\begin{proposition}\label{Lax pair}
An explicit Lax pair for the Hamiltonian form of any Painlev\'e
equation, leads an explicit Lax pair for the Hamiltonian form of the
corresponding conjugate Painlev\'e equation. The relevant
construction involves the following steps:
\begin{enumerate}
\item [(i)] Substitute the $t-$function into the Lax pair of a given Painlev\'e equation, so that the new independent variables become $\la$ and $h$ (instead of $\la$ and $t$).
\item [(ii)] Replace in the resulting Lax pair the unknown functions by the associated explicit functions of $(p,q,h)$.
\end{enumerate}

\end{proposition}
\paragraph{Proof.}
Let $H(p,q,t)$ be a Hamiltonian of a given Painlev\'e equation, i.e., the given Painlev\'e equation is equivalent to
Hamilton's equations
\begin{equation}
\label{Hamilton} \f{dp}{dt}=-\f{\pa H}{\pa q},~~~\f{dq}{dt}=\f{\pa
H}{\pa p}.
\end{equation}
Let $T(p,q,h)$ be the associated conjugate Hamiltonian, i.e., the associated
conjugate Painlev\'e equation is equivalent to Hamilton's
equations
\begin{equation}
\label{CHamilton} \f{dp}{dh}=\f{\pa T}{\pa q},~~~\f{dq}{dh}=-\f{\pa
T}{\pa p}.
\end{equation}
Suppose equations \eqref{Hamilton} admit the following Lax pair:
\begin{subequations}\label{LAX pair general HAM}
\begin{align}
&\f{\pa \psi}{\pa \la}(\la,t)=A(p(t),q(t),t,\la)\psi(\la,t),\\
&\f{\pa \psi}{\pa t}(\la,t)=B(p(t),q(t),t,\la)\psi(\la,t),~~~~~~~~\la\in\mathbb{C},
\end{align}
\end{subequations}
where $A$ and $B$ are two known $k\times k$ matrix-valued functions
of $(p,q,t)$, and the function $\psi$ is a $k\times k$
matrix-valued function of $\la$ and $t$ (for some $k>1$). Equations \eqref{LAX pair general HAM} imply Lax's equation
\begin{equation} \label{LAX_General}
\pa_t A-\pa_{\la}B+[A,B]=0, \footnote{We mention that $\pa_t
A=\f{\pa A}{\pa p}\f{dp}{dt}+\f{\pa A}{\pa q}\f{dq}{dt}+\f{\pa
A}{\pa t}$,}
\end{equation}
where $[~\cdot~,~\cdot~]$ denotes the usual matrix commutator.

Let $h=h(t)$ denote the $h-$function and let $t=t(h)$ denote the
$t-$function, which is the inverse of the $h-$function. Let
$\phi(\la,h)=\psi(\la,t(h))$. Replacing in equations \eqref{LAX pair general HAM} $\psi$ by $\phi$, we find
\begin{subequations}\label{LAX pair general trans}
\begin{align}
&\f{\pa \phi}{\pa \la}(\la,h)=A(p(t(h)),q(t(h)),t(h),\la)\phi(\la,h),\\
&\f{\pa \phi}{\pa h}(\la,h)=\f{dt}{dh}
B(p(t(h)),q(t(h)),t(h),\la)\phi(\la,h).
\end{align}
\end{subequations}
Both the $h-$function and the $t-$function are unknown
functions (the knowledge of these functions requires solving
Hamilton's equations). However, by the definition of the
conjugate Hamiltonian, we have
\begin{equation}\label{a}
t(h)=T(p(t(h)),q(t(h)),h).
\end{equation}
Moreover, the conjugate Hamiltonian structure \eqref{CHamilton}
implies
\begin{equation}\label{b}
\f{dt}{dh}=\f{\pa T}{\pa h}.
\end{equation}
Using in equations \eqref{LAX pair general trans} equations \eqref{a} and
\eqref{b} to replace the unknown functions in terms of explicit
functions, we obtain the following Lax pair for equations
\eqref{CHamilton}:
\begin{subequations}\label{LAX pair final}
\begin{align}
&\f{\pa \phi}{\pa \la}=A(p,q,T(p,q,h),\la)\phi,\\
&\f{\pa \phi}{\pa h}=\f{\pa T}{\pa h} B(p,q,T(p,q,h),\la)\phi.
\end{align}
\end{subequations}
Indeed, Lax's equation reads:
\begin{equation}\label{LAX_Gen_C}
\pa_h A-\f{\pa T}{\pa h} \pa_{\la}B + \f{\pa T}{\pa h} [A, B]=0.
\end{equation}
Noting that $$\pa_h=\f{dt}{dh}\pa_t =\f{\pa T}{\pa h} \pa_t,$$ we
find that equation \eqref{LAX_Gen_C} is just equation
\eqref{LAX_General} using $h$ as the independent variable.
\begin{flushright}$\square$\end{flushright}

\begin{example} (A Lax pair for $CP_I$ and $\widetilde{CP_I}$)
Recall the Jimbo-Miwa pair \cite{JM}\cite{Kita} for Hamilton's equations \eqref{Ham_P1} of Painlev\'e I:
\begin{subequations}\label{Lax_P1_JM}
 \begin{align}
\f{\pa \psi}{\pa \la}&=\left\{\left(
\begin{array}{llcl}
-p  & q^2+t/2\\
-4q & p\\
\end{array}
\right)+\la\left(
\begin{array}{llcl}
0 & q\\
4 & 0\\
\end{array}
\right)+\la^2\left(
\begin{array}{llcl}
 0 & 1\\
 0 & 0\\
\end{array}
\right)\right\}
\psi,\\
\f{\pa \psi}{\pa t}&=\left\{\left(
\begin{array}{llcl}
0 & q\\
2 & 0\\
\end{array}
\right)+\la\left(
\begin{array}{llcl}
 0 & 1/2\\
 0 & 0\\
\end{array}
\right)\right\}
\psi.
 \end{align}
\end{subequations}
Replacing in equations \eqref{Lax_P1_JM} $t$ by $$\f{1}{2}\f{p^2}{q}-2q^2-\f{h}{q}$$ and using
$$\f{\pa T}{\pa h}=-\f{1}{q},$$
we obtain the following Lax pair for $CP_I$ and $\widetilde{CP_I}$:
\begin{subequations}\label{Lax_CP1_JM}
 \begin{align}
\f{\pa \psi}{\pa \la}&=\left\{\left(
\begin{array}{llcl}
-p  & \f{p^2}{4q^2}-\f{h}{2q}\\
-4q & p\\
\end{array}
\right)+\la\left(
\begin{array}{llcl}
0 & q\\
4 & 0\\
\end{array}
\right)+\la^2\left(
\begin{array}{llcl}
 0 & 1\\
 0 & 0\\
\end{array}
\right)\right\}
\psi,\\
\f{\pa \psi}{\pa h}&=-\f{1}{q}\left\{\left(
\begin{array}{llcl}
0 & q\\
2 & 0\\
\end{array}
\right)+\la\left(
\begin{array}{llcl}
 0 & 1/2\\
 0 & 0\\
\end{array}
\right)\right\}
\psi.
 \end{align}
\end{subequations}
This Lax pair can be verified directly.
\end{example}
\section{Conclusions}
We have introduced a novel class of integrable ODEs, which are
related to $P_{I}$, $P_{I\!I}$, $P_{I\!V}$ and $\widetilde{P_{I}}$,
$\widetilde{P_{I\!I}}$ and $\widetilde{P_{I\!V}}$. The relation
between the new ODEs and the Painlev\'e equations is
\textit{implicit}. We recall, that there exist analogous implicit
relations among integrable PDEs, namely the relations derived via the
so-called hodograph transformations. For example, the celebrated
Korteweg-de Vries and Harry-Dym equations are related by precisely
such a transformation \cite{CFA}.

Hodograph type transformation do \textit{not} preserve the
Painlev\'e property (for example, solutions for the Harry-Dym
equation do \textit{not} possess this property \cite{CFA}).
Similarly, we do not expect that conjugate Painlev\'e equations to
possess the Painlev\'e property. Nevertheless, these equations
\textit{are} integrable. Indeed, it is possible to construct a large
class of solutions of the conjugate equations. Furthermore, in
principle, it is possible to express the solution of the general
initial value problem in terms of the solutions of the initial value
problem of the associated Painlev\'e equation. However, the most
efficient way to solve the initial value problem of a given
conjugate ODE, is to use its associated Lax pair. For the conjugate
equations of $P_{I\!I}$ and $P_{I}$, relevant Lax pairs are given by
equations \eqref{Lax_CP2} and \eqref{Lax_CP1_JM}. For other
conjugate ODEs, similar Lax pairs can be constructed using
Proposition \eqref{Lax pair}.

Taking into consideration the relation between the implicit
transformations discussed here and hodograph type transformations,
it is natural to expect that the ODEs introduced here might appear
as ODE reductions of integrable PDEs (such as the Harry-Dym
equation), which are related to well known integrable PDEs (such as
the Korteweg-de Vries equation) via hodograph transformations.

\paragraph{Acknowledgements}
D. Yang would like to thank Professor Youjin Zhang for his advise
and for helpful discussions, as well as the China Scholarship
Council for supporting him for a joint PhD study at the University
of Cambridge. A. S. Fokas is grateful to the Guggenheim Foundation,
USA, for partial support.

\bibliographystyle{amsplain}

\end{document}